\documentclass[prx,aps,amsmath,amssymb,twocolumn,showpacs,superscriptaddress,10pt]{revtex4-2}

\usepackage{graphicx}
\usepackage{bm}
\usepackage{bbold}

\usepackage{nicematrix}
\usepackage[colorlinks,bookmarks=true,citecolor=blue,linkcolor=blue,urlcolor=blue]{hyperref}
\allowdisplaybreaks
\usepackage{siunitx}
\usepackage{subfigure}
\usepackage{soul}

\usepackage[dvipsnames]{xcolor}

\newcommand{\tr}{\textrm{tr}}
\newcommand{\upp}{\mathrm{p}}
\newcommand{\ups}{\mathrm{s}}
\newcommand{\ii}{\mathrm{i}}

\DeclareSIUnit\sample{Sa}

\makeatletter
\renewcommand*{\fnum@figure}{{\normalfont\bfseries \figurename~\thefigure}}
\makeatother

\begin{document}

\title{Multi-dimensional parameter space of higher-order exceptional points induced by Brillouin optoacoustics}

\author{Grigorii Slinkov}
\email{grigorii.slinkov@mpl.mpg.de}
\affiliation{Max Planck Institute for the Science of Light, 91058 Erlangen, Germany}
\affiliation{Department of Physics, Friedrich-Alexander-Universit\"at Erlangen-N\"urnberg, 91058 Erlangen, Germany}
\author{Anton Montag}
\affiliation{Max Planck Institute for the Science of Light, 91058 Erlangen, Germany}
\affiliation{Department of Physics, Friedrich-Alexander-Universit\"at Erlangen-N\"urnberg, 91058 Erlangen, Germany}
\author{Julius T. Gohsrich}
\affiliation{Max Planck Institute for the Science of Light, 91058 Erlangen, Germany}
\affiliation{Department of Physics, Friedrich-Alexander-Universit\"at Erlangen-N\"urnberg, 91058 Erlangen, Germany}
\author{Quentin Levoy}
\affiliation{Max Planck Institute for the Science of Light, 91058 Erlangen, Germany}
\affiliation{Laboratoire Collisions Agrégats Réactivité, Université de Toulouse, CNRS, 31062 Toulouse, France}
\author{Paulina Fuentes Rivera}
\affiliation{Max Planck Institute for the Science of Light, 91058 Erlangen, Germany}
\affiliation{Department of Physics, Friedrich-Alexander-Universit\"at Erlangen-N\"urnberg, 91058 Erlangen, Germany}

\author{Flore K. Kunst}
\email{flore.kunst@mpl.mpg.de}
\affiliation{Max Planck Institute for the Science of Light, 91058 Erlangen, Germany}
\affiliation{Department of Physics, Friedrich-Alexander-Universit\"at Erlangen-N\"urnberg, 91058 Erlangen, Germany}
\author{Birgit Stiller}
\email{birgit.stiller@mpl.mpg.de}
\affiliation{Max Planck Institute for the Science of Light, 91058 Erlangen, Germany}
\affiliation{Department of Physics, Friedrich-Alexander-Universit\"at Erlangen-N\"urnberg, 91058 Erlangen, Germany}
\affiliation{Institute of Photonics, Leibniz University Hannover, 30167 Hannover, Germany}
\date{\today}

\begin{abstract}
    Exceptional points (EPs) are degeneracies in the spectrum of non-Hermitian systems, where both the eigenvalues and eigenvectors coalesce.
    In the vicinity of an $n\mathrm{th}$ order EP, the eigenvalues generally show $n\mathrm{th}$-root dependence on the system parameters, making EPs potentially promising candidates for ultra-sensitive measurements.
    Usually EPs are implemented in precisely fabricated nano- and microstructures.
    In this work, we instead show the experimental implementation of a third-order EP (EP3) using the synthetic dimension in a single-mode optical fiber, leveraging multi-frequency Brillouin scattering.
    We perform a multi-dimensional scan of the parameter space revealing not only an EP3 but also additional topological structures connected to it.
    Our work paves the way toward fabrication-free realizations of exceptional points of arbitrary order.
\end{abstract}

\maketitle

\section*{Introduction}

\begin{figure*}[th]
    \centering
        \includegraphics[width=.99\linewidth]{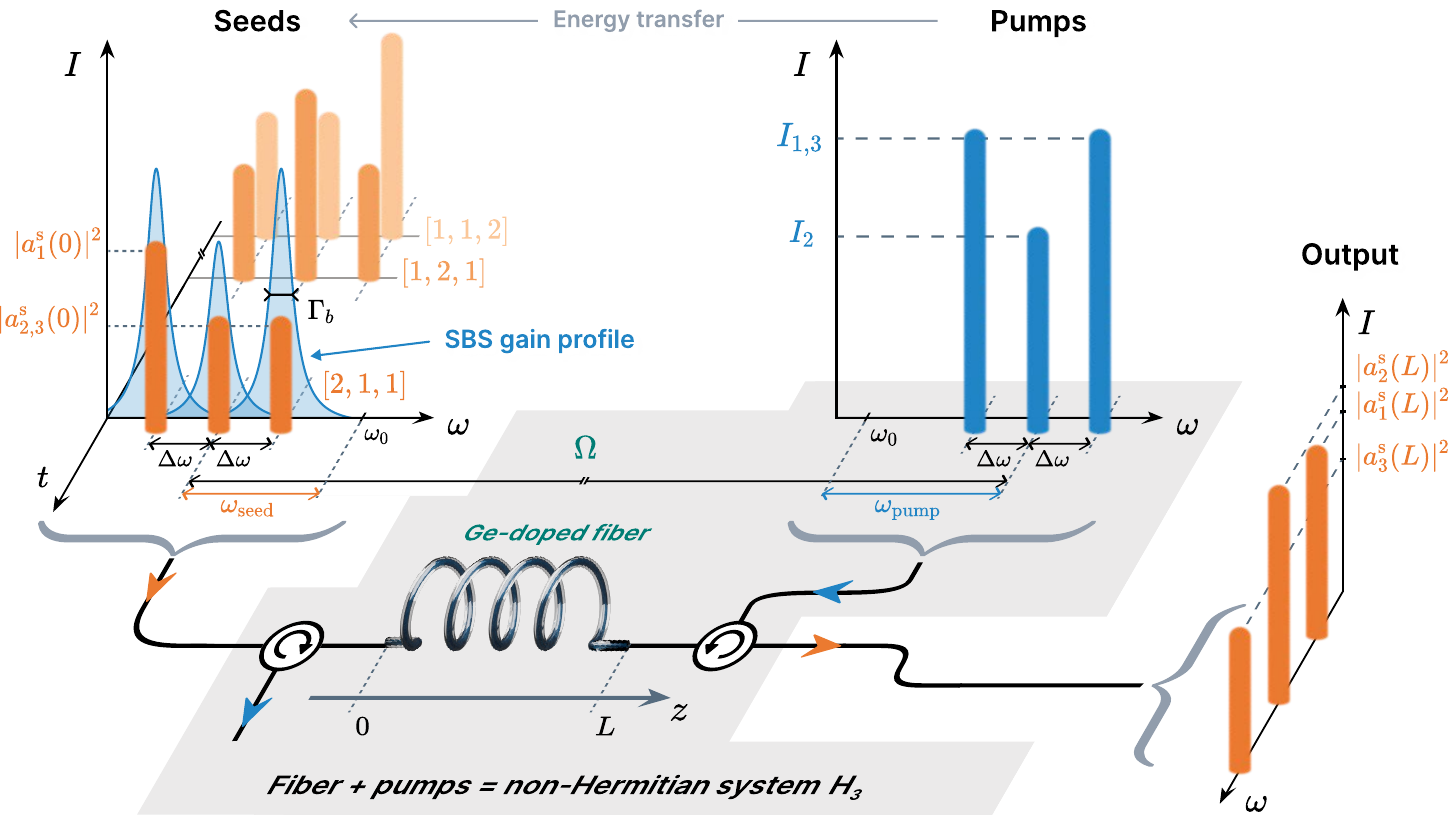}
    \caption{
    \textbf{Conceptual scheme of the experiment.}
    The non-Hermitian dynamical matrix $H_3$ describes the evolution of the three seeds (orange) traveling through a system (highlighted in gray) comprising a Germanium-doped single-mode fiber and three pumps (blue) injected into it.  
    The intensities of the pumps $I_{2}$, $I_1=I_3=I_{1,3}$, and the frequency spacing $\Delta\omega$ between them span the three-dimensional parameter space wherein the EP3s and accompanying topological structures lie.
    In order to reveal these features, the seeds are sent into the system, counter-propagating with the pumps.
    Each seed-pump pair is matched to the Brillouin frequency of the Ge-doped optical fiber~$\Omega$.
    These seeds and pumps are created from the same laser source with frequency $\omega_0$ via electro-optic modulation; the seeds are downshifted in frequency and the pumps are upshifted (see Sec.~\ref{sec: setup},~\ref{sec: pump tones}).
    The Brillouin interaction between the seeds and the pumps takes place over the length $L$ of the doped fiber with the seed propagating in the positive $z$-direction.
    This process transfers energy from the pumps to the seeds and creates coupling between the seeds.
    The system is probed by sending in a given seed configuration $\vec{a^\ups}(0) = [a_1^\ups(0),a_2^\ups(0),a_3^\ups(0)]^\mathrm{T}$.
    The input and the output seed configurations are connected via the dynamical matrix $H_3$ of the system as $\vec{a^\ups}(L) = e^{-\ii H_3L}\vec{a^\ups}(0)$. 
    Measuring the transmission for three linearly independent seed configurations 
    (labeled by their amplitude ratios $[1,1,2]$, $[1,2,1]$ and $[2,1,1]$) allows us to reconstruct the dynamical matrix of the system and extract its eigenvalues and eigenvectors.
    The behavior of the eigenvalues and eigenvectors upon sweeping the system parameters reveals the presence of the EP3s.
    }
    \label{fig: graph_abst}
\end{figure*}

Eigenvalues and eigenvectors are instrumental in the description of physical systems. 
The evolution of classical systems can be decomposed into a set of elementary excitations encoded with eigenvectors, while eigenvalues represent the corresponding resonant frequencies.
In certain non-conservative (non-Hermitian) systems, exceptional points (EPs) can be found, where a unique phenomenon occurs: the eigenvectors coalesce, resulting in the breakdown of the eigenmode decomposition.
The exotic behavior enabled by this breakdown has drawn increased interest to EPs -- especially, the prospect of highly sensitive measurement. 
While this is a subject of heated academic discussions~\cite{Wiersig2014,Wiersig2016,Langbein2018,Lau2018,Zhang2019,Chen2019,Wiersig2020,Wiersig2022,Duggan2022,Wiersig2026}, the experimental implementation of EPs motivated researchers from various fields to challenge their devices, fabrication methods and measurement schemes to proof the existence of these features. 
Examples of platforms for these studies are microwave billiards~\cite{Dembowski2001,Dembowski2004}, acoustic gratings~\cite{Fang2021} and acoustic cavities~\cite{Tang2023}, semiconducting monolayers~\cite{Johns2025}, electrical circuits~\cite{Yin2023,Suntharalingam2023,Bai2024,Li2024a}, trapped ions~\cite{Chen2025}, Rubidium vapors~\cite{Niu2024}, and Bose-Einstein condensates~\cite{Wang2024}.
There has also been a plethora of exotic applications, ranging from coherent control of magnon-polaritons~\cite{Lambert2025} and enhanced magnon-based Brillouin light scattering~\cite{Liang2025} to bistable memory~\cite{Chen2022}, large-scale unidirectional transparency~\cite{Feng2013}, negative refraction~\cite{Fleury2014}, mode switching~\cite{Liu2022}, quantum state transfer~\cite{Zhang2025} and many more.

\begin{figure*}[th]
    \centering
    \includegraphics[width=0.99\linewidth]{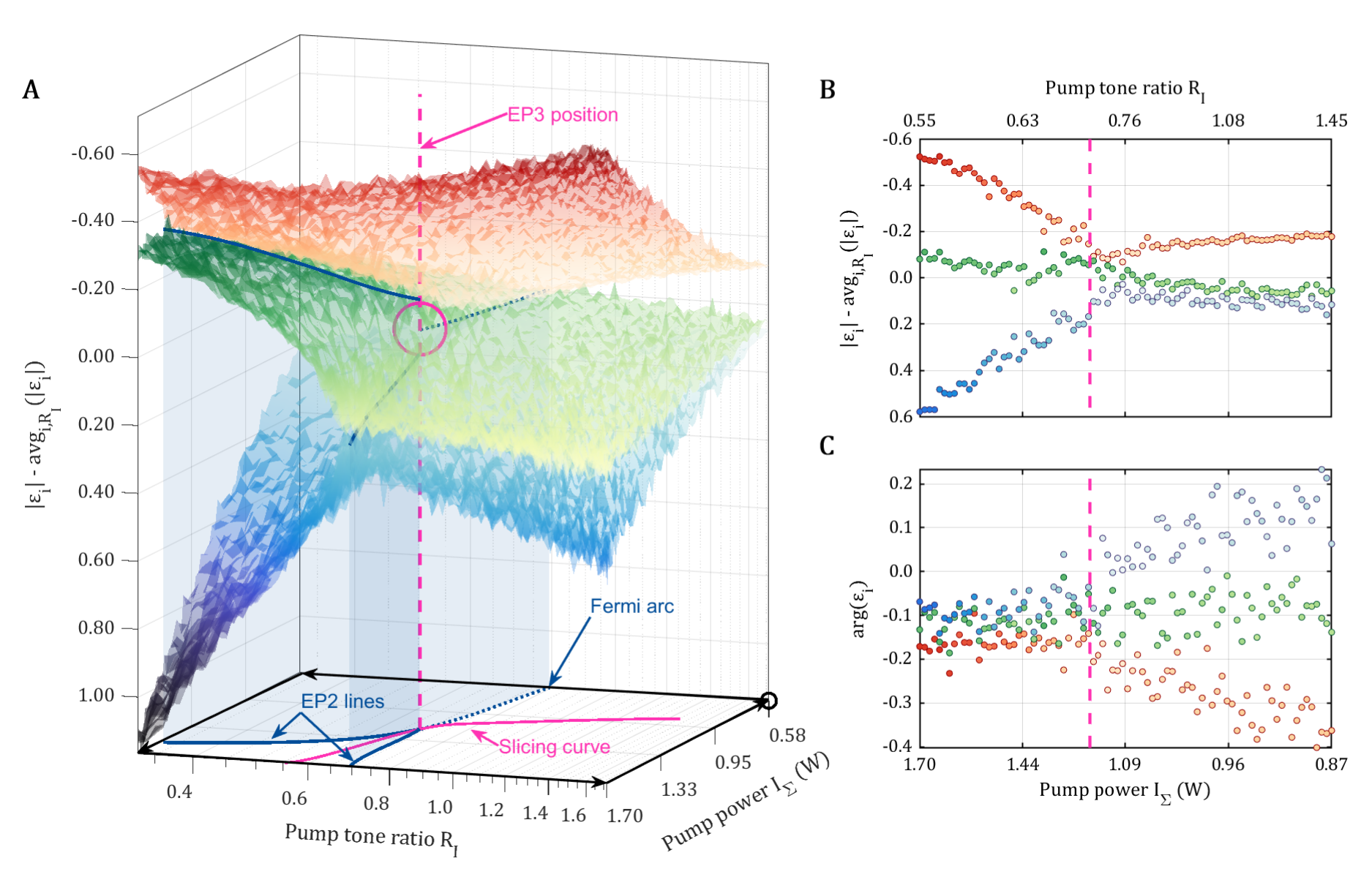}
    \caption{
    \textbf{Exceptional points of the transmission matrix.}
    Panel~\textbf{A} shows three surfaces, each corresponding to the magnitude of one of the three eigenvalues of the non-Hermitian transmission matrix $T=e^{-\ii H_3L}$.
    The plot is obtained at a fixed frequency spacing $\Delta \omega$ by varying the intensities $I_2$ and $I_1=I_3$ of the three pumps.
    We present the results using pump power $I_\Sigma = I_1+I_2+I_3$ and pump tone ratio $R_I = I_{1,3}/I_2$ coordinates.
    To improve the readability of the plot we subtract $\textrm{avg}_{i,R_I}(|\varepsilon_i|)$, the mean eigenvalue magnitude averaged over the pump tone ratio, from $|\varepsilon_i|$. 
    This mean corresponds to the average gain experienced by the seeds
    The position of the EP3 in the two-dimensional parameter space is marked by a vertical dashed line (magenta); the corresponding spectral degeneracy is indicated by a circle in the center of the plot (magenta).
    The positions of two EP2 lines (solid blue) and an imaginary Fermi arc (dotted blue) originating at the EP3 are marked in the parameter space at the bottom.
    These lines are additionally projected up onto the eigenvalue surfaces to highlight their respective spectral features.
    A black circle in the far right corner marks the start of the encircling loop used in Fig.~\ref{fig: loop}, with black arrows indicating the loop. 
    Panels~\textbf{B} and \textbf{C} are obtained by slicing the dataset along the solid magenta curve marked at the bottom of the plot in panel~\textbf{A}.
    The position of the EP3 is again marked with a dashed magenta line.
    In panel~\textbf{B} we show again the magnitudes of the eigenvalues, whereas in~\textbf{C} we additionally show their phases.
    Left of the EP3, all magnitudes are different and the phases equal, whereas right of the EP3 two magnitudes are equal and all three phases are different, as predicted for this slice~\cite{Montag2024,Montag2026}.
    }
    \label{fig: EP3_3D}
\end{figure*}

Photonics has been a fruitful playground for the experimental investigation of EPs due to the ability to engineer gain and loss, tuning non-Hermiticity with great precision.  
Furthermore, photonics offers possibilities to couple different modes, increasing the accessible order of an EP, as its order (the number of coalescing eigenvectors) is limited by the number of interacting modes.
In particular, the simplest EPs -- EPs of second order (EP2s) -- have been implemented in photonic systems such as dielectric mirrors~\cite{Feng2013}, photonic molecule lases~\cite{Brandstetter2014}, photonic crystals~\cite{Zhen2015,Zhou2018}, microcavities~\cite{Peng2016,Chen2022}, optical chip waveguides~\cite{Liu2022,Schumer2022}, and free-space cavities~\cite{Ruan2025}.

However, implementing EPs of higher order has proven difficult, because it requires precise control of an increased number of system parameters.
Only few realizations of higher-order EPs in optical systems have been experimentally demonstrated. 
Among them are ring resonators~\cite{Hodaei2017,Jahangiri2025}, a cavity optomechanical system~\cite{Patil2022} and an experimental simulation enabled by single-photon interferometry~\cite{Wang2023}.

Due to the vastness of the enlarged parameter space, finding higher-order EPs demands a possibility of wide multi-parameter sweeps.
Fabricated structures usually only allow narrow parameter sweeps (e.g., via heating) without refabrication.
Recently, a fabricationless realization of an EP2 has been demonstrated in an optical fiber based on stimulated Brillouin-Mandelstam scattering (SBS)~\cite{bergman2021}.
SBS is a nonlinear effect that couples optical and traveling acoustic waves in bulk materials, resonators and optical waveguides.
While it is a nuisance for telecommunications \cite{Ippen1972}, it has also has been advantageously used for lasers~\cite{gundavarapu_sub-hertz_2019, otterstrom_silicon_2018, chauhan_visible_2021}, sensing~\cite{galindez-jamioy_brillouin_2012, geilen_extreme_2023}, gyroscopes~\cite{li_microresonator_2017, lai_earth_2020}, microscopy~\cite{antonacci_recent_2020, prevedel_brillouin_2019, scarcelli_confocal_2008}, microwave photonics ~\cite{marpaung_integrated_2019}, and most recently, neuromorphic optoacoustic processing~\cite{Zeng2023,Geilen2023,Becker2024,Slinkov2025,Merklein2017,Zhu2007}.

In order to show the existence of EPs in any non-Hermitian system, one has to ideally map out the parameter space surrounding the EP.
Previous experiments were limited in terms of multi-dimensional measurements and therefore had to probe the dispersion along selected lines in parameter space.
Furthermore, access to eigenvectors is often very restricted in experimental realizations, for which reason only partial measurements can be carried out to validate the theory.

In this work, we go beyond these limitations: we measure both eigenvalues and eigenvectors within the multidimensional parameter space and provide a high-resolution scan of the eigenvalue surfaces. We experimentally demonstrate the existence of an EP3 and reveal topological structures originating at the EP3: two lines of EP2s and a bulk Fermi-arc~\cite{Montag2024,Kozii2017}, along which we find a partial degeneracy of the eigenvalues. 
The experimental realization is based on multi-frequency SBS in a commercially available single-mode optical fiber. 
It is validated by our theoretical model in which on- and off-resonant frequency channels form a non-Hermitian system, which gives access to higher-order exceptional points and accompanying topological features. 
This fabrication-free approach replaces the design and fabrication of complicated micro- and nano-structures with the generation of optical signals to manipulate coherent light states via volatile traveling acoustic waves.
Our versatile implementation offers reproducible and fast creation and detection of all-optically controlled EP3s in standard optical fibers. 
As this implementation is possible to be implemented on a photonic chip, it can be combined with other topological structures. 
Our work demonstrates that it is possible to resolve parameter spaces containing higher-order topological features with high resolution, which would require almost impossible fabrication accuracy if implemented in a manufactured device.

\subsection*{SBS as source of gain and mode coupling}

Stimulated Brillouin scattering is a third-order nonlinear effect that couples a pair of counter-propagating optical waves with a traveling acoustic wave serving as a mediator between them~\cite{wolff_brillouin_2021}. 
It follows the phase-matching condition~\cite{stiller_cross_2019}: the frequencies of the optical waves propagating in opposite directions have to be separated by the acoustic wave's frequency $\Omega$, which is defined by the optical wavelength and the properties of the interaction medium. 
In the Stokes process, the energy is transferred from a higher frequency wave (pump) to a lower frequency wave (seed).
For such a single seed-pump pair of waves the process can be described by 
\begin{align}
        \ii\partial_z a^\ups &= \ii g_0\mathcal{P}_0\frac{|A^\upp|^2}{1+\ii\Gamma} a^\ups, &
        \Gamma &= \frac{\Omega^2-(\omega^\upp-\omega^\ups)^2}{\Omega\Gamma_b}   ,
        \label{eq: seed-pump x1}
\end{align}
where $a^\ups$ is the complex amplitude of the seed wave with frequency $\omega^\ups$ that propagates in the positive $z$-direction, $g_0$ is the Brillouin gain coefficient (in \unit[]{\per \meter \per \watt}), $\mathcal{P}_0|A^\upp|^2$ is the intensity (in \unit[]{\watt}) of the pump wave with frequency $\omega^\upp$ and dimensionless complex amplitude $A^\upp$, and $1/\Gamma_b$ is the acoustic lifetime.
Equation~\eqref{eq: seed-pump x1} is obtained using an undepleted pump and a local acoustic response approximations~\cite{wolff_brillouin_2021}.
These are well justified given the assumptions of a strong pump (order of magnitude higher than the seed), continuous wave (CW) light (or slowly varying envelope) and a long waveguide (with respect to acoustic wave propagation distance).
It is evident from Eq.~\eqref{eq: seed-pump x1} that the seed experiences amplification, which grows with the pump intensity and waveguide length. 
Tuning away from the resonant case $\omega^\upp-\omega^\ups=\Omega$ decreases the amplification, quantified by $\Gamma$.
We will refer to this as off-resonant SBS.

The picture complicates if we introduce a second seed-pump pair, and assume both pairs to be perfectly matched, i.e., $\omega^\upp_{1,2}-\omega^\ups_{1,2}=\Omega$.
If we bring the two pairs sufficiently close to each other $\omega^{\upp,\ups}_2-\omega^{\upp,\ups}_1 \sim \Gamma_b$, each seed will draw energy from the other pump off-resonantly.
Furthermore, bringing the two seed-pump pairs this close allows us to assume that the interaction between the seed and the pump in one pair is mediated by the same acoustic wave that mediates the interaction of the second pair~\cite{bergman2021}.
This way, SBS does not only provide gain, but also introduces a mechanism to couple different frequency-multiplexed optical modes.
This principle can be extended to an arbitrary combination of pump and seed modes, which lays the foundation of the present work and is developed into a generalized theory of multi-frequency off-resonant Brillouin scattering in Ref.~\citenum{Montag2026}.

\section*{Results}

\begin{figure*}[th]
    \centering
    \includegraphics[width=\linewidth]{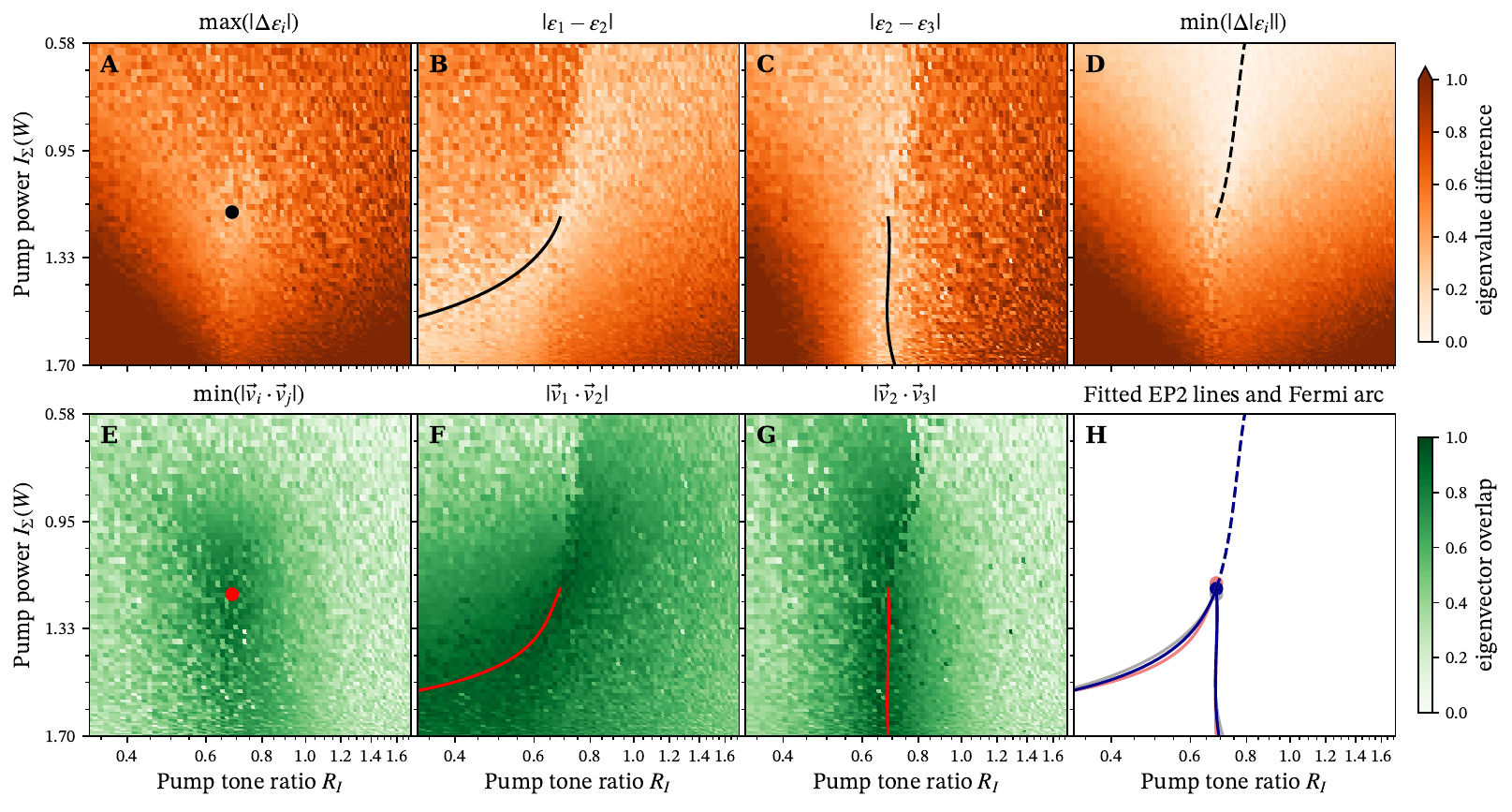}
    \caption{
    \textbf{Eigenvalue differences and eigenvector overlaps.} 
    Panels \textbf{A}-\textbf{D} show the distance between the eigenvalues~$\varepsilon_{1,2,3}$. 
    Panel \textbf{A} shows the maximal distance between all three complex eigenvalues. 
    The EP3 (black dot) is located at the minimum of the data presented in this plot.
    Panels \textbf{B} and \textbf{C} show two pairwise eigenvalue differences between the two lowermost and two uppermost sheets in Fig.~\ref{fig: EP3_3D}, respectively. 
    The EP2 lines (solid black lines) are extracted by tracing out the valley originating at the EP3.
    Panel \textbf{D} shows the minimal difference between the eigenvalue magnitudes.
    This plot is used to find the position of the imaginary Fermi arc (dashed black line).
    Panels \textbf{E}-\textbf{G} show the overlaps (absolute value of the pairwise scalar product) between the eigenvectors $\vec{v}_{1,2,3}$. 
    Panel \textbf{E} shows the minimal of the three overlaps.
    The EP3 is given by the peak in this plot (red dot).
    Panels \textbf{F} and \textbf{G} show two pairwise overlaps corresponding to the two lowermost and two uppermost sheets in Fig.~\ref{fig: EP3_3D}, respectively. 
    The EP2 lines (solid red lines) are extracted by tracing out the ridges in the overlaps starting at the EP3.
    Panel \textbf{H} shows all extracted features from panels \textbf{A}-\textbf{D} in gray and from panels \textbf{E}-\textbf{G} in red, with the averaged EP3 position and EP2 lines shown in blue. 
    }
    \label{fig: overlap}
\end{figure*}

In order to demonstrate the capability to induce high-order EPs, we choose a configuration with three pump waves with frequencies~$\omega^\upp_{1,2,3}$ and three seed waves with frequencies~$\omega^\ups_{1,2,3}$, which are sent from the opposite ends into an optical fiber as shown in Fig.~\ref{fig: graph_abst}. 
This choice allows us to observe an EP3, as the maximal order of an EP is tied to the number of seeds.
Each seed-pump pair is perfectly matched and the individual pairs are spaced by $\Delta\omega$, so we have
\begin{subequations}
\label{eq: freqs}
\begin{align}
        \omega^\upp_{1,2,3}-\omega^\ups_{1,2,3}&=\Omega, \label{eq:bigomega}\\
        \omega^{\upp,\ups}_3-\omega^{\upp,\ups}_2 &=\Delta\omega = \omega^{\upp,\ups}_2-\omega^{\upp,\ups}_1. \label{eq:deltaomega}
\end{align}
\end{subequations}
If the spacing between seed-pump pairs is sufficiently small ($\Delta\omega\sim\Gamma_b)$, the shared acoustic wave creates coupling between them.
Furthermore, the close spacing necessitates taking off-resonant interactions into account.
Applying the undepleted (stationary) pump approximation allows us to only consider the coupling that is created between the seeds. 
As the set of three seeds $\vec{a^\ups}=[a^\ups_1,a^\ups_2,a^\ups_3]^\mathrm{T}$ propagates through the fiber in the positive $z$-direction, its evolution is governed by a Schrödinger-like equation
\begin{equation}
\label{eq: schroedinger}
    \ii\partial_z\vec{a^\ups} = H_3\vec{a^\ups},
\end{equation}
where the $3\times3$ dynamical matrix $H_3$ (the full form is provided in Sec.~\ref{sec: the matrix}) describes both the coupling between the seeds and the gain they experience.
With help of Eq.~\eqref{eq: schroedinger}, an EP3 can be found by observing how this non-Hermitian matrix affects the seeds upon sweeping the parameters of the system.
In order to facilitate this search, we reduce the dimensionality of the parameter space by imposing anti-parity-time (anti-$\mathcal{PT}$) symmetry on $H_3$ as detailed in Sec.~\ref{sec: symmetry}.
This operation preserves the EP3, but leaves only three parameters to tune: The intensities of the outer pump tones $I_{1,3} = \mathcal{P}_0|A^\upp_{1,3}|^2$, the intensity of the middle pump tone $I_{2} = \mathcal{P}_0|A^\upp_{2}|^2$, and the frequency spacing $\Delta\omega$, which is reflected in the conceptual scheme of the experiment in Fig.~\ref{fig: graph_abst}.

Since the dynamical matrix $H_3$ does not depend on the seeds Eq.~\eqref{eq: schroedinger} can be solved for a known fiber length $L$ as
\begin{align}
    \label{eq:T}
    \vec{a^\ups}(L) = T \cdot \vec{a^\ups}(0), \quad\textrm{with}\;\;  T = e^{-\ii H_3L}.
\end{align}
With help of this simple relation, the transmission matrix~$T$ can be probed by comparing the initial state $\vec{a^\ups}(0)$ with the state after the transmission through the system $\vec{a^\ups}(L)$. 
Probing with three linearly independent seed configurations allows us to fully reconstruct~$T$ (details of this procedure are presented in Sec.~\ref{sec: tomography}).
The topological features of $T$ are carried over from $H_3$ such that the two matrices can be used interchangeably.
Indeed, while both matrices share the same eigenvectors, the imaginary parts of eigenvalues of $H_3$ correspond to the magnitudes of the eigenvalues of $T$, and the real parts of eigenvalues of $H_3$ correspond to the phases of the eigenvalues of $T$.
The magnitude of the eigenvalues of $T$ represent the accumulated gain the seeds experience.
These considerations combined with the fact that the transmission matrix is accessed directly in the experiment justify the presentation of the experimental findings using the eigenvalues and the eigenvectors of the transmission matrix $T$.

In Fig.~\ref{fig: EP3_3D}~\textbf{A} we present what we believe to be the most captivating result of this paper - the spectrum (its absolute part) of the transmission matrix $T$ measured experimentally revealing the EP3.
The measurement is performed sweeping a two-dimensional parameter space for a fixed frequency spacing $\Delta\omega/2\pi = \SI{80}{\mega\hertz}$. 
Each of the surfaces corresponds to one of the three eigenvalue magnitudes, sorted by magnitude and colored for readability.
The following anti-$\mathcal{PT}$ symmetry-induced topological structures are visible in the plot:
\begin{itemize}
    \item The EP3 (magenta circle) is found where all three eigenvalue sheets coalesce to one point;
\end{itemize}
and originating from it
\begin{itemize}
    \item the lower EP2 line (blue solid line) is found where the bottom (blue) and the middle (green) sheets merge;
    \item the upper EP2 line (blue solid line) is found middle (green) and top (red) sheets merge;
    \item the Fermi arc (blue dotted line) is found where all three sheets intersect.
\end{itemize}

The parameter sweep in Fig.~\ref{fig: EP3_3D} (as well as in the subsequent plots) is performed over the intensities of the pumps.
The choice of coordinates -- the ``pump tone ratio" $R_I = I_{1,3}/I_2$ and the ``pump power" $I_\Sigma = I_1+I_2+I_3$ -- reflects the specificity of the measurement setup, which is described in Sec.~\ref{sec: pump tones}.
As the pumps are generated via electro-optical modulation, their intensities can be precisely adjusted and reconfigured (see Sec.~\ref{sec: pump tones}).
This trait allows us to not only explore the parameter space freely to find the EP3, but also to cover a very wide parameter space around it with fine resolution to show the accompanying symmetry-induced topological features in unprecedented detail.

In order to study the eigenvalue behavior in the EP3 vicinity more closely, we plot the eigenvalue magnitudes and the phases found along a slice cutting through the EP3 separately in Fig.~\ref{fig: EP3_3D}~\textbf{B,C} with a shared horizontal axis.
The slice is curved in order to avoid overlap with the EP2 lines and the imaginary Fermi arc to showcase qualitative changes in the eigenvalues when passing through the EP3.
Between the EP2 lines all magnitudes are different and the phases equal. 
After crossing the EP3 the two magnitudes are equal while all three phases are different, as predicted for such a slice~\cite{Montag2024,Montag2026}.

To gain a deeper insight into the topological features, we present a series of plots showing eigenvalue differences and eigenvector overlaps in Fig.~\ref{fig: overlap}.
From these plots we extract the positions of the topological features.
Panel \textbf{A} shows the maximal distance between the eigenvalues, which reaches its minimum at the EP3, where all three eigenvalues coalesce.
Along the EP2 lines two of the eigenvalues coalesce, which we employ to trace the EP2 lines through the parameter space as shown in panels \textbf{B} and \textbf{C}.
When plotting the minimal difference of the eigenvalue magnitudes, c.f. panel \textbf{D}, the position of the imaginary Fermi arc is found by tracing along the valley originating at the EP3.
A major advantage of our experimental platform is that we can fully reconstruct the transmission matrix, gaining access not only to the eigenvalues, but to the eigenvectors as well.
This enables a complementary approach for finding the topological features.
The minimal overlap of the eigenvectors is maximal at the EP3, because there all three eigenvectors coalesce, c.f. panel \textbf{E}.
The pairwise overlaps $|\vec{v}_1\cdot\vec{v}_2|$ and $|\vec{v}_2\cdot\vec{v}_3|$ are maximal along the EP2 lines, along which the corresponding eigenvectors coalesce.
We highlight this in panels \textbf{F} and \textbf{G}, where we determine the EP2 line positions from the overlaps.
In panel \textbf{H} the fitted EP3, EP2 line and the Fermi arc positions determined from the eigenvalue difference (gray) and the eigenvector overlap (red) are compared and the averaged positions of these features are shown (blue). 
We notice that the positions of the various topological features extracted using these two methods are in good agreement with each other.
Note that the position of the Fermi arc can only be extracted from the eigenvalues and not the eigenvectors.
These averaged features are the ones displayed in Fig.~\ref{fig: EP3_3D}.

\begin{figure}[b]
    \centering
    \includegraphics[width=\linewidth]{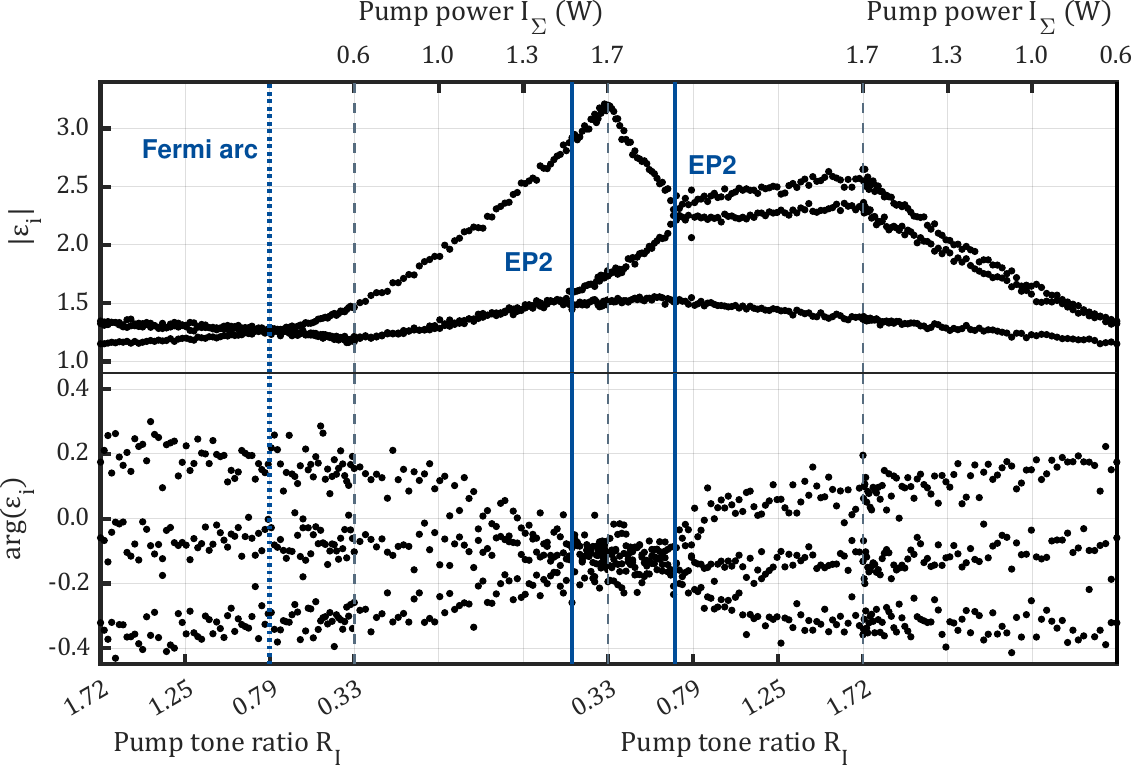}
    \caption{\textbf{Encircling the EP3.}
    Magnitudes (top) and phases (bottom) of the three eigenvalues are plotted following a loop that encircles the EP3. 
    The sides of the loop are formed by intermittent segments of ``constant pump tone ratio, variable pump power" and ``constant pump power, variable pump tone ratio", and coincide with the boundaries of the swept area in Fig.~\ref{fig: EP3_3D}~\textbf{A}. 
    The corners of the swept area are marked with vertical dashed gray lines.
    The dotted vertical blue line marks the point where the loop crosses the imaginary Fermi arc and the two solid vertical blue lines where the loop crosses the lower EP2 line (left) and the upper EP2 line (right).
    The presence of these three spectral features -- the imaginary Fermi arc, the lower EP2 and the upper EP2 -- prove the existence of the EP3 inside the encircled area.
    }
    \label{fig: loop}
\end{figure}

Finally, we provide the definitive proof of the presence of the EP3: Figure~\ref{fig: loop} displays the eigenvalues measured along a closed loop around the EP3, which follows the boundaries of the swept area as shown in Fig~\ref{fig: EP3_3D}~\textbf{A}. 
Along this loop a threefold degeneracy of the magnitude of the eigenvalues, marking the imaginary Fermi arc, and two distinct EP2s are clearly visible. 
The EP2s appear once for the two eigenvalues with the larger magnitude (upper EP2) and once for those with the smaller magnitude (lower EP2). 
Finding these three topological features along a closed loop is proof that a symmetry-induced EP3 is encircled~\cite{Montag2024,Montag2026}.
In fact, we used this criterion to initially find the EP3 in the experiment: A loop was moved around in the parameter space until all three telltale features were found along the loop. 
This search process is described in detail in Sec.~\ref{sec: encircling}.

\section*{Discussion}

In this work we have experimentally realized a higher-order exceptional point in a fabrication-free photonic system. 
The versatility of our approach has allowed us to measure the eigenvalues and eigenvectors in the neighborhood of a symmetry-induced EP3, revealing the rich topological structures that accompany it.

At the core of our approach lies multi-frequency off-resonant stimulated Brillouin scattering that creates non-Hermiticity and establishes mode coupling at the same time.
SBS allows us to all-optically control light (seeds) using light (pumps) mediated by acoustic waves, and thus equips our approach with high-speed and high-bandwidth capabilities for applications requiring fast and efficient information processing.
The fact that Brillouin scattering can also be implemented in optical chips~\cite{eggleton_brillouin_2019, Merklein2017, choudhary_advanced_2017, morrison_compact_2017}
offers a pathway for miniaturization and integration with other photonic devices, further expanding the potential applications of our approach. 

Adding to the potential of Brillouin scattering, a number of advantages come from using optical fibers as platform.
Choosing the frequency domain to be the host dimension for the interacting optical modes allowed us to use a single-mode single-core Germanium-doped fiber.
As this fiber is directly compatible with standard telecom single-mode fibers, it allows integration of non-Hermitian phenomena into modern-day photonic instruments.
As an example, the enhanced sensitivity offered by EPs can be readily harnessed to improve the performance of well-established SBS applications, for instance, strain and temperature sensing.
Furthermore, employing SBS in an optical waveguide, such as our fiber, does not limit us to modes of resonator modes of any kind, granting additional freedom and simplifying operation. 

A critical building block enabling our findings is the use of electro-optical modulation to control the pump composition (and thus the dynamical matrix). 
In this way, the weight of fine-tuning the system parameters is not carried by precise nanofabrication. 
It is instead lifted by the ease of generating electrical signals with their unmatched configurability and reproducibility.
This capability proved to be particularly beneficial in the context of EPs, where small changes of parameters can lead to dramatic changes in the behavior of the system.

Finally, the theoretical proposal that inspired this work covers EPs of arbitrary order, and even further describes $N$-seed-$M$-pump off-resonant multi-frequency Brillouin scattering~\cite{Montag2026}.
This generalization gives researchers the tool to construct a wide variety of non-Hermitian systems.
The experimental setup presented here allows us to test the theoretical predictions that follow in a highly controlled, but at the same time remarkably flexible environment.
This combination of theory and experiment offers a powerful toolbox to design and experimentally realize intricate non-Hermitian systems.

\section*{Materials and Methods} \label{sec: Methods}

\subsection{The dynamical matrix} \label{sec: the matrix}

The non-Hermitian dynamical matrix $H_3$ introduced in Eq.~\eqref{eq: schroedinger} describes the coupling between the three seeds and the gain they experience propagating through the fiber.
It is obtained under the experimentally met assumptions of undepleted (stationary) pumps, the continuous wave approximation and a local acoustic response. 
While the full derivation can be found in Ref.~\citenum{Montag2026}, the final result is
\begin{widetext}
    \begin{equation}\label{eq:33effH}
        H_3 = \ii g_0 \mathcal{P}_0 \begin{pmatrix}
         |A_1^\upp|^2 + \frac{|A_2^\upp|^2}{1+\ii \gamma}+ \frac{|A_3^\upp|^2}{1+2 \ii \gamma} - \frac{\ii \Gamma_b \gamma}{c g_0 \mathcal{P}_0} & A_1^\upp (A_2^\upp)^* + \frac{A_2^\upp (A_3^\upp)^*}{1+\ii \gamma} & A_1^\upp (A_3^\upp)^* \\
        A_2^\upp (A_1^\upp)^* + \frac{A_3^\upp (A_2^\upp)^*}{1+\ii \gamma} & \frac{|A_1^\upp|^2}{1-\ii \gamma} + |A_2^\upp|^2 + \frac{|A_3^\upp|^2}{1+\ii \gamma} & A_2^\upp (A_3^\upp)^* + \frac{A_1^\upp (A_2^\upp)^*}{1-\ii \gamma} \\
        A_3^\upp (A_1^\upp)^* & A_3^\upp (A_2^\upp)^* + \frac{A_2^\upp (A_1^\upp)^*}{1-\ii \gamma}  & \frac{|A_1^\upp|^2}{1-2\ii \gamma} + \frac{|A_2^\upp|^2}{1-\ii \gamma} + |A_3^\upp|^2 + \frac{\ii \Gamma_b \gamma}{c g_0 \mathcal{P}_0}
    \end{pmatrix}  - 2k_0 \, \mathbb{1}_3 \, .
    \end{equation}
\end{widetext} 
Here, $g_0$ is the Brillouin gain coefficient, $\mathcal{P}_0$ is the power density connecting dimensionfull pump intensities $I_{1,2,3} = \mathcal{P}_0|A^\upp_{1,2,3}|^2$ with dimensionless complex-valued pump amplitudes $A^\upp_{1,2,3}$, $\Gamma_b$ is the Brillouin linewidth, $c=c_0/n$ is the speed of light in the fiber with $c_0$ the speed of light in vacuum and $n$ the effective refractive index, and $k_0=\omega^\ups_2/c$ accounts for the propagation of the seeds through the interaction region.
As each seed is chosen to be strictly resonant with its designated pump, $\gamma$ is used to quantify the detuning with the off-resonant pumps $\gamma = 2\Delta\omega/\Gamma_{b}$.

The qualitative considerations presented in the introduction find their reflection in the elements of the dynamical matrix. 
The imaginary parts of the diagonal elements describe the gains individual seeds experience passing through the interacting region.
They include the gain provided by each pump, accounting for off-resonant interactions by detuning factors. 
On the other hand the real parts of the diagonal elements (including the term $- 2k_0 \mathbb{1}_3$) describe the phase evolution of the seeds.
The off-diagonal elements describe coupling between different seeds, and correspondingly, only contain products of different pump amplitudes.

The dynamical matrix $H_3$ contains two sets of parameters. 
One is defined by the fiber sample: The Brillouin gain coefficient $g_0$, the Brillouin frequency $\Omega$ (cf. Eq.~\eqref{eq:bigomega}), and the Brillouin linewidth $\Gamma_b$.
The other set is determined by the light going into the fiber: The three complex-valued pump amplitudes $A^\upp_{1,2,3}$ and the frequency spacing $\Delta\omega$ (cf. Eq.~\eqref{eq:deltaomega}).
This parameter set can be continuously tuned during the experiment, facilitating the search for EP3s.

Importantly, Eq.~\eqref{eq: schroedinger} with $H_3$ from Eq.~\eqref{eq:33effH} describes a linear evolution of the seed amplitudes given the assumptions presented in Ref.~\citenum{Montag2026}.
This allows us to view the combination of the fiber and the pump light as a composite non-Hermitian system, as illustrated in Fig.~\ref{fig: graph_abst}. 
From this description we obtain the transmission matrix $T$ according to Eq.~\eqref{eq:T}, and the EP3 is found by analyzing its eigenvalues.
From this perspective, the sole role of the seeds is to probe the transmission of the system and thus convey the EP3 presence.

\subsection{
\texorpdfstring{Imposing anti-$\mathcal{PT}$ symmetry}{Imposing anti-PT symmetry}} \label{sec: symmetry}

As discussed in Sec.~\ref{sec: the matrix} the set of experimentally tunable parameters contains the three complex pump amplitudes~$A^\upp_{1,2,3}=|A^\upp_{1,2,3}|e^{\ii \phi^\upp_{1,2,3}}$ and the frequency spacing~$\Delta\omega$. 
Thus, the parameter space is spanned by seven real-valued coordinates, the three pump magnitudes $|A^\upp_{1,2,3}|$, the three pump phases $\phi^\upp_{1,2,3}$ and $\Delta\omega$.
This is sufficient for the realization of an EP3, which requires $2(n-1)|_{n=3}=4$ real constraints to be satisfied~\cite{Delplace2021, Sayyad2022}. 
However, the seven-dimensional parameter space is vast and especially having to control the pump phases is experimentally challenging.
In order to reduce both the parameter space dimension and the number of real constraints on the EP3, we impose anti-$\mathcal{PT}$ symmetry~\cite{Montag2024}. 
Therefore we restrict the traceless part of the dynamical matrix $\Tilde{H}_3=H_3-\left[\tr(H_3)/3\right]\mathbb{1}_3$ by the symmetry condition
\begin{align}
    \Tilde{H}_3 = - \Theta\Tilde{H}_3^*\Theta^{-1}, \quad \textrm{with} \;     \Theta = \begin{pmatrix}
        0&0&1\\
        0&1&0\\
        1&0&0
    \end{pmatrix} \, .
\end{align}
This leads to a pair of conditions on the pump magnitude $|A^\upp_{1,2,3}|$ and phases $\phi^\upp_{1,2,3}$
\begin{subequations}
\begin{gather}
|A^\upp_1| = |A^\upp_3|, \\
    \phi^\upp_1-2\phi^\upp_2+\phi^\upp_3=0 .\label{eq:contphi}
\end{gather}
\end{subequations}
As long as the condition in Eq.~\eqref{eq:contphi} is met, the occurrence of the EP3 is independent of the specific pump-phase values~\cite{Montag2026}.
In particular, Eq.~\eqref{eq:contphi} is always fulfilled if the initial phase configuration created by the electro-optic modulation (see Sec.~\ref{sec: setup}) is set to $\phi^\upp_1=\phi^\upp_2=\phi^\upp_3$.
While the free evolution of the pump phases throughout the fiber changes each phase, it does not violate Eq.~\eqref{eq:contphi} due to the symmetric frequency spacing.
Therefore setting the initial phases as $\phi^\upp_1=\phi^\upp_2=\phi^\upp_3$ reduces the parameter space dimension to three: The outer pump magnitudes $|A^\upp_1| = |A^\upp_3|$, the central pump magnitude $|A^\upp_2|$, and the frequency spacing $\Delta \omega$.

\subsection{Transmission matrix tomography} \label{sec: tomography}

As stated before, the transmission matrix $T=e^{-\ii H_3L}$ can be reconstructed by measuring the state of the seed before (at $z=0$) and after the interaction (at~$z=L)$.
Each such measurement gives three linear complex-valued equations 
\begin{align*}
    \begin{pmatrix}
        a_1^\ups\\
        a_2^\ups\\
        a_3^\ups
    \end{pmatrix}_{z=L} \, =
    T\cdot
    \begin{pmatrix}
        a_1^\ups\\
        a_2^\ups\\
        a_3^\ups
    \end{pmatrix}_{z=0},
\end{align*}
while the matrix contains nine unknown complex elements.
Therefore, one has to perform the measurement for a set of three linearly independent seed configurations to determine the full matrix.
This is reflected in Fig.~\ref{fig: graph_abst}, where we show three seed configurations being sent into the system sequentially. 
We label them as $[1,1,2]$, $[1,2,1]$ and $[2,1,1]$, which corresponds to the ratios of the amplitudes used in the experiment.
Combining the three column vectors corresponding to these seed configurations into matrices as
\begin{subequations}
\begin{align}
\label{eq: M(L)}
    M(L) &=
    \begin{pmatrix}
        \vec{a^\ups}_{112} & \vec{a^\ups}_{121}  & \vec{a^\ups}_{211}
    \end{pmatrix}_{z=L},
\\
\label{eq: M(0)}
    M(0) &=
    \begin{pmatrix}
        \vec{a^\ups}_{112} & \vec{a^\ups}_{121}  & \vec{a^\ups}_{211}
    \end{pmatrix}_{z=0},
\end{align}
\end{subequations}
allows us to write a simple expression for $T$:
\begin{align}
\label{eq: T from Ms}
    T = M(L)M(0)^{-1}.
\end{align}
Once the matrix is reconstructed, the eigenvalues and eigenvectors are extracted numerically. 
This routine is repeated for each point of the parameter  sweep.

\subsection{EP3 search routine (encircling)} \label{sec: encircling}

Exceptional points reveal themselves in the eigenvalue behavior when the system parameters are swept. 
Applying the anti-$\mathcal{PT}$ symmetry, as discussed in Sec.~\ref{sec: symmetry}, allows us to reduce the dimensionality of the parameter space. 
This leaves three coordinates: the intensities of outer pump tones 
$I_1 = I_3$,
the intensity of the middle tone 
$I_2$
and the frequency spacing $\Delta\omega$.
It is shown in Ref.~\citenum{Montag2026} that in this three-dimensional space the EP3s form a line at the cusp of an EP2 surface where it meets with a bulk Fermi surface.
Fixing one of the three parameters is equivalent to slicing this volume with a plane.
Intersecting the EP3 line gives a single point and intersecting the EP2 and Fermi surfaces results in lines that meet at the EP3. 
This way the EP3 can be found by sweeping either two of the three parameters.
More so, the symmetry-induced structures help find it.

The symmetry-induced EP3 can be found by measuring eigenvalues along closed loops in the parameter space (encircling). 
When such a loop crosses one of the special lines (an EP2 line or a Fermi arc), the crossing point will be revealed in the behaviour of the eigenvalues. 
It is has been shown that all three lines (an upper EP2 line, a lower EP2 line and a Fermi arc) start at the EP3~\cite{Montag2024}. 
This way, a loop that crosses each of the three lines exactly once is guaranteed to contain an EP3 inside it.

If we slice the $[I_{1,3},I_2,\Delta\omega]$ parameter space by a fixed frequency spacing plane, the Fermi arc originating at the EP3 will continue in the direction of lower pump power, while the EP2 lines that also start at the EP3 will generally continue in the direction of higher pump power. 
This way, to find the EP3 it is convenient to either find a Fermi arc and gradually increase power or find an EP2 line and gradually decrease it.

\subsection{Experimental setup} \label{sec: setup}

\begin{figure}[tbh]
\centering
\includegraphics[width=.99\linewidth]{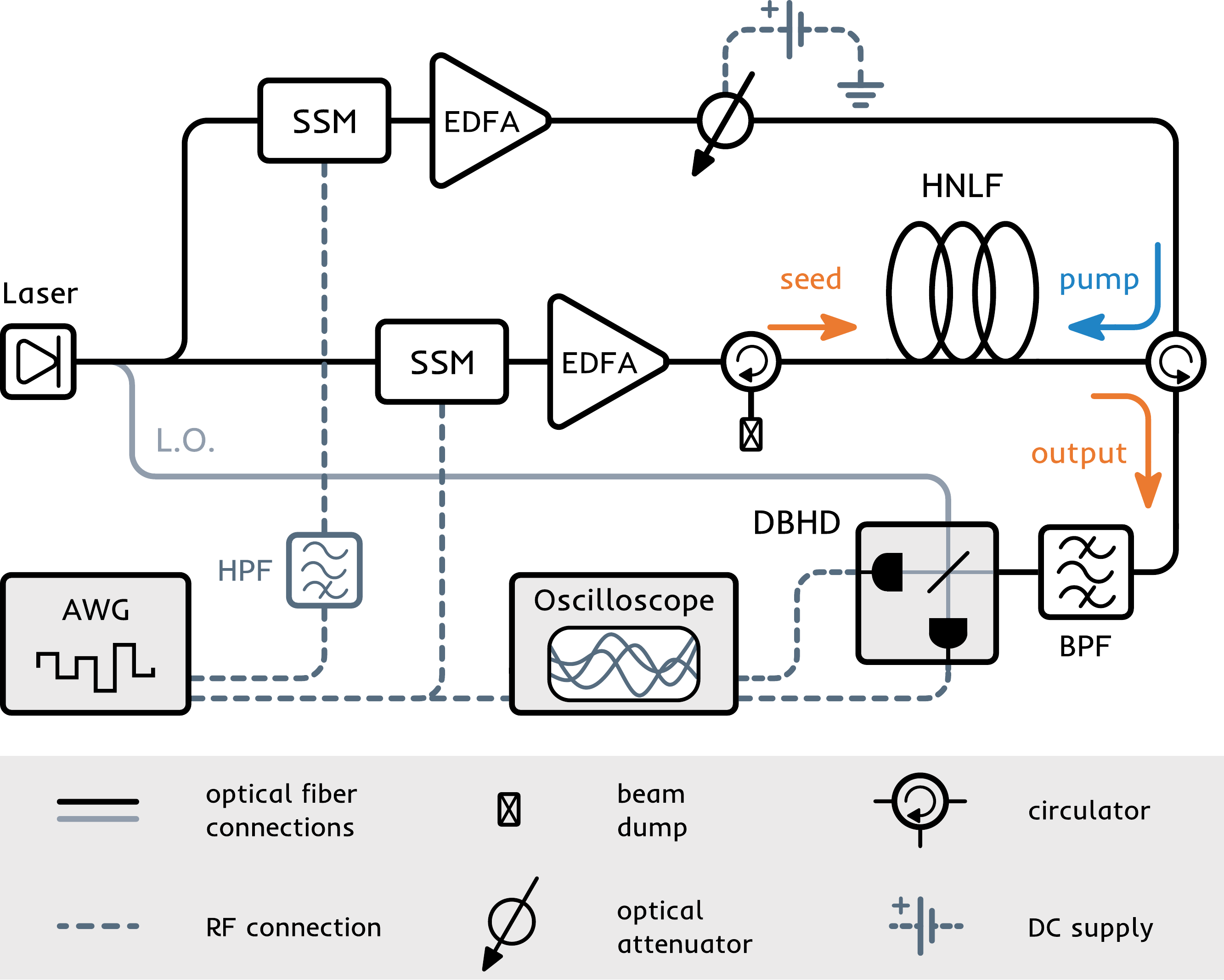}
\caption{\textbf{Experimental setup scheme.} The following abbreviations are used: L.O. - local oscillator, SSM -- single sideband modulator, EDFA -- Erbium-doped fiber amplifier, HNLF -- highly nonlinear fiber, BPF -- optical bandpass filter, DBHD -- double balanced heterodyne detector, HPF -- high-pass RF filter, AWG -- arbitrary waveform generator.}
\label{fig: exp_scheme}
\end{figure}

Navigating the parameter space requires versatile control over a set of system parameters: frequency spacing $\Delta\omega$, middle tone intensity $I_2$, and the intensity of the side tones $I_{1,3}$.
This control has been achieved using a setup depicted schematically in Fig.~\ref{fig: exp_scheme}. 
The output of a single CW laser source with frequency $\omega_0/2\pi=\SI{193.41}{\tera\hertz}$ (which we will refer to as carrier) is split to create both the pumps and the seeds.
Each of the two arms features an electro-optical single sideband modulator (SSM), driven with one of the two channels of an arbitrary waveform generator (AWG). 
Each channel feeds a composite three-tone radio frequency (RF) signal to the respective SSM, generating the required seeds and pumps.

Once the optical signals are generated, the pump arm is amplified with a two-stage erbium-doped fiber amplifier (EDFA). 
The combined power of the pump tones can be lowered using the subsequent variable attenuator that is controlled electronically. This combined power serves as one of the coordinates for the experimental measurement (see Sec.~\ref{sec: pump tones}).
The seed is amplified using a pre-amplifier EDFA that provides linear gain. 
The pump and the seed then interact inside a single-mode Germanium-doped highly nonlinear fiber (HNLF). 
The parameters of the fiber are: gain coefficient $g_0 = \SI{1.7}{\per\meter\per\watt}$, Brillouin linewidth $1/(2\pi\Gamma_B) = \SI{45}{\mega\hertz}$, Brillouin frequency $\Omega/2\pi = \SI{9.73}{\giga\hertz}$~\cite{Geilen2024} and length $L=\SI{20}\meter$.

After the interaction the pumps are directed into a beam dump and the seeds are sent to the double balanced homodyne detector (DBHD) together with the local oscillator.
Before detection the output light is additionally filtered with an optical bandpass filter (BPF) to further suppress the original carrier and the unwanted sideband generated by the SSM.
The data is collected using an oscilloscope. 
The phase-sensitive detection allows us to obtain both the amplitude and the relative phases of each of the three constituent seeds.

\subsection{Tone generation} 
\label{sec: waveforms}
The three monochromatic optical waves (tones) for the pump and the seed arms are created using electro-optic modulation. 
The pump and the seed SSMs are fed with their respective RF signals generated by a designated AWG channel:
\begin{subequations} \label{eq: RF_tones}
    \begin{align}
        &\text{seed channel} & U^\ups(t) &= \sum_{i=1}^3 U^\ups_i\cos{(2\pi f^\ups_i t)}, \\
        &\text{pump channel} & U^\upp(t) &= \sum_{i=1}^3 U^\upp_i\cos{(2\pi f^\upp_i t)}.
    \end{align}
\end{subequations}
Each harmonic component of such a signal creates a pair of optical sidebands: one shifted up  from the laser frequency to $\omega_i^{\ups,\upp} = \omega_0+2\pi f^{\ups,\upp}_i$ 
and the other one shifted down to $\omega_i^{\ups,\upp} = \omega_0-2\pi f^{\ups,\upp}_i$.
The biases of the SSM can be adjusted to suppress either of these two sidebands.
The intensities of the sidebands are proportional to the amplitude of the driving signal $U^{\ups,\upp}_i$.
This way, electro-optic modulation gives precise and versatile control over the individual amplitudes of the seed and pump tones and their frequencies.

SBS requires a fixed frequency shift between the pump and the seed, defined by the fiber in use: 
$\omega^\upp_i - \omega^\ups_i = \Omega = 2\pi \cdot \SI{9.73}{\giga\hertz}$. 
We split this difference between the two channels by using the higher-frequency sideband for the pump arm (suppressing the carrier and the lower sideband) and the lower-frequency sideband for the seed arm (suppressing the carrier and the higher sideband).
This way, the requirement for the modulation frequencies becomes $\Omega/2\pi = f^\upp_i+f^\ups_i$.

For the seed arm we choose the base frequency (the frequency of the middle tone, indicated in Fig.~\ref{fig: graph_abst} as $\omega_\textrm{seed}$) to be $f^\ups_2 = \SI{2}{\giga\hertz}$. 
The two other tones are spaced equally to the sides of it (see Fig~\ref{fig: graph_abst}): $f_{1,3}^\ups = f_2^\ups\pm \Delta f$ with the spacing $\Delta f = \Delta\omega/2\pi$ lying in the tens of megahertz range. 
This fits the \SI{2.5}{\giga\hertz} bandwidth of the DBHD setup but at the same time allows us to use an optical bandpass filter to further suppress the carrier.

On the pump arm side this choice forces a $f^\upp_2=\SI{7.73}{\giga\hertz}$ shift onto the middle tone of the pump (indicated in Fig.~\ref{fig: graph_abst} as $\omega_\textrm{pump})$.
Choosing a relatively high base frequency for the seed allows us to decrease the base frequency of the pump.
Still, the desired frequency comes out higher than half the sampling rate $\textrm{SR} = \SI{12}{\giga\sample\per\second}$ of the AWG  used in the experiment.
We circumvent this issue by using the second Nyquist zone image of the signal in combination with RF filtering. 
In reality, an AWG tasked to produce a single harmonic signal additionally generates a whole family of aliases. 
Each next frequency tone carries less power than the last, but they can nevertheless be used if a higher frequency-signal is needed.
The position of the second alias (first one being the original signal) is a mirror image of the original with the reflecting plane positioned at half of the sampling rate.
This way, generating the second alias at the desired frequency $f^\upp_{\textrm{alias}_2} = \SI{7.73}{\giga\hertz}$ requires the pump channel RF frequency to be $f^\upp_2=\textrm{SR}/2 +(\textrm{SR}/2-f^\upp_{\textrm{alias}_2}$) = \SI{4.27}{\giga\hertz}.
Applying a high-pass RF filter (HPF) blocks the original frequency but lets through the high-frequency alias, allowing us to effectively generate a signal that exceeds half the sampling rate of the AWG.

\subsection{Controlling the pump intensities}
\label{sec: pump tones}

As discussed in Sec.~\ref{sec: encircling}, sweeping only two of the three free parameters -- frequency spacing, side tone intensity and middle tone intensity -- is sufficient to locate the EP3.
Even though using an AWG to generate the pump tones allows us to control both the frequencies and the complex amplitudes of the pumps, in the experiment it was decided to fix the frequency spacing $\Delta\omega$.
The reason is that the interplay between the changing pump frequencies and the fixed sampling rate of the AWG reflects in the magnitudes of the generated tones, making it hard to control them independently. 

Furthermore, the two remaining coordinates $I_{1,3}$ and $I_2$ were transformed in accordance with a specificity introduced by the pump arm EDFA. 
While providing high output power, this double-stage EDFA also tends to output a fixed amount of energy with no regard to the composition of its input.
This means that, for example, decreasing the magnitude of one of the pump tones via decreasing the driving voltage of that tone would increase the magnitudes of the two remaining tones.
This issue was mitigated by switching to a new set of coordinates that is invariant with regard to this interdependency.

The first coordinate is chosen to be the combined intensity of the three pump tones $I_\Sigma=I_1+I_2+I_3$.
Facilitating this change, we employ a voltage-controlled optical attenuator at the output of the EDFA (Fig.~\ref{fig: exp_scheme}) as it provides  faster and more precise control than the native EDFA function.

The second adjustment concerns the composition of the pump waveforms. 
The goal is to vary the pump tone ratio $R_I=I_{1,3}/I_2$ in a way that would keep the total power $I_\Sigma$ constant, making the two coordinate axes orthogonal.
In order to achieve that, as we adjust the amplitude of the harmonic component that generates the middle tone $U^\upp_2$ (see Eq.~\eqref{eq: RF_tones}), we simultaneously scale the other two, keeping their sum $U^\upp_2+2U^\upp_{1,3}$ constant. 
Although the tone voltages do not translate perfectly into the tone intensities due to nonlinear responses of the SSM and the EDFA, this scaling still serves the purpose of sweeping the parameter space more efficiently. 

These coordinates, which we refer to as pump power~($I_\Sigma$) and pump tone ratio ($R_I$), are used throughout the paper to present the experimental findings.
It is easy to see that the two coordinate systems $[I_{1,3},I_2]$ and 
$[I_\Sigma,R_I]$
can be used interchangeably as long as neither of the pump intensities is zero.

\subsection{Adjusting for free propagation} \label{sec: free_prop}

The transmission matrix reconstruction routine presented in Sec.~\ref{sec: tomography} requires us to measure the seeds at the start and the end of the HNLF.
The real-world implementation requires two additional considerations.

First, instead of measuring the seeds $M(0)$ at the start of the fiber (see Eq.~\eqref{eq: M(0)}), we measure the transmitted seeds at the output of the system when the pumps are turned off.
In this case the propagation of the seeds through the HNLF Eq.~\eqref{eq:T} is governed not by $T$, but by a unitary free propagation matrix
\begin{align}
    U&=\begin{pmatrix}
        \label{eq:free_prop}
        \exp(\ii \Delta k L) & 0 & 0 \\ 0 & 1 & 0 \\ 0 & 0 & \exp(-\ii \Delta k L)
    \end{pmatrix},
\end{align}    
where $L$ is the fiber length, $\Delta k = \Delta\omega \, n/c$, $n$ is the effective refractive index of the fiber, and  $c$ is the speed of light.

Second, the setup includes a section where the seeds propagate freely from end of the HNLF to the DBHD scheme (see Fig.~\ref{fig: exp_scheme}). 
This section cannot be eliminated in the experiment, but can be accounted for using an additional unitary free propagation matrix $U_\textrm{out}$, which has the same form as Eq.~\eqref{eq:free_prop}.

Combining these two considerations, we can write down expressions for the seeds measured in the experiment:
\begin{subequations}
    \begin{align}
    \label{eq: total_Msa}
        M_\textrm{off} &= U_\text{out} U M(0), \\
        M_\textrm{on} &= U_\text{out} T M(0).\label{eq: total_Msb}
    \end{align}
\end{subequations}
To determine the eigenvalues and eigenvectors of the transmission matrix, we evaluate
\begin{equation}
    \text{eig}(T) = \text{eig}(M_\textrm{on}M_\textrm{off}^{-1}U)\, .
\end{equation}
This can be shown by inserting Eqs.~\eqref{eq: total_Msa} and \eqref{eq: total_Msb}.
By switching the order of the diagonal unitaries $U_\textrm{out}^{-1}$ and $U$ we obtain
\begin{align*}
    &\text{eig}(M_\textrm{on}M_\textrm{off}^{-1}U) \\
    &= \text{eig}\left(U_\text{out} T M(0)(M(0))^{-1}U^{-1}U_\textrm{out}^{-1}U\right) \\
    &= \text{eig}(U_\text{out} T U^{-1}UU_\textrm{out}^{-1}) \\
    &= \text{eig}(U_\text{out} T U_\textrm{out}^{-1})\\
    &= \text{eig}(T) 
\end{align*}
because the unitary $U_\text{out}$ does not affect the eigenvalues and the eigenvectors.

\subsection{Eigenvalue extraction routine} \label{sec: eig extract}

At the output of the setup, the seed light is detected using the DBHD scheme and the signal is recorded using the oscilloscope (Fig.~\ref{fig: exp_scheme}). 
The two traces containing the I and the Q quadratures are combined into one complex-valued array, which contains information about both the magnitudes and the phases of the tones.
For each of the three seed configurations (e.g., $[1,1,2]$, $[1,2,1]$, and $[2,1,1]$) the seed waveform is played continuously in a loop for the duration of a measurement.
The pump waveform is gated with a \SI{50}{\percent} duty cycle. 
This allows us to capture both the ``pump off" and the ``pump on" cases (introduced in Sec.~\ref{sec: free_prop}) in a single oscilloscope trace.
This trace is then split according to these cases and each of the two segments is fitted with a function of form 
$\sum_{i=1}^{3}(X_i+\ii Y_i)\cos{(2\pi f^\ups_i \cdot t)}$. 
Note that only $X_i$ and $Y_i$ are the fit parameters here, the frequencies of the tones are known apriori.

As the three starting configurations are measured separately (unlike the ``SBS on" and ``SBS off" cases, which are obtained in a single oscilloscope trace), the phase fluctuates significantly between these measurements. 
In order to account for that, we rotate each of the three ``on-off" pairs independently such that the phase of the middle tone ($\omega^\ups_2$) of the ``SBS off" measurement becomes zero.
This effectively puts the three configurations into the same reference frame, allowing us to exclude the effect of phase noise.

\section*{Acknowledgments}
The authors thank Andreas Geilen and Gladys Jara-Schultz for help with the setup and visualization.

\subsection*{Funding} 
G.S., P.F.R. and B.S. acknowledge funding from the European Unions’s ERC Consolidator Grant ”Sound-PC” (101170362), the Deutsche Forschungsge-
meinschaft (DFG) under grants STI-792/7-1 and STI-
792/1-1, and the Max Planck Society through the Inde-
pendent Max Planck Research Groups scheme. A.M., J.T.G., and F.K.K. acknowledge funding from the Max Planck Society Lise Meitner Excellence Program~\mbox{2.0}. A.M., J.T.G. and F.K.K. also acknowledge support from the European Union’s ERC Starting Grant “NTopQuant” (101116680). The views expressed are those of the authors and do not necessarily reflect those of the European Union or the ERC.

\subsection*{Author contributions} 
Initial idea: BS, FK, QL; Conceptualization:  GS, QL, AM, JG, FK, BS; Methodology: GS, QL, AM, JG, BS; Investigation: GS, PFR; Experiment: GS;
Visualization: GS, AM; Supervision: FK, BS; Writing — original draft: GS, BS

\subsection*{Competing interest} The authors declare no competing interests.

\subsection*{Data Availability}

All data supporting the findings of this study is available from the corresponding author upon reasonable request.

\bibliography{Optoacoustic_NAF_.bib,EP3_arxiv.bib}

\end{document}